# Analysis of Relationship between Strategic and Aggregate Energy Minimization in Delay-Constrained Wireless Networks

Rajgopal Kannan and Shuangqing Wei     Louisiana State University

*Abstract*— We formulate two versions of the power control problem for wireless networks with latency constraints arising from duty cycle allocations In the first version, strategic power optimization, wireless nodes are modeled as rational agents in a power game, who strategically adjust their powers to minimize their own energy. In the other version, joint power optimization, wireless nodes jointly minimize the aggregate energy expenditure. Our analysis of these models yields insights into the different energy outcomes of strategic versus joint power optimization. We derive analytical solutions for power allocation under both models and study how they are affected by data loads and channel quality. We derive simple necessary conditions for the existence of Nash equilibria in the power game and also provide numerical examples of optimal power allocation under both models. Finally, we show that joint optimization can (sometimes) be Pareto-optimal and dominate strategic optimization, i.e the energy expenditure of all nodes is lower than if they were using strategic optimization.

## I. INTRODUCTION

Energy-efficiency is a critical concern in many wireless networks, such as cellular networks, ad-hoc networks or wireless sensor networks (WSNs) that consist of large number of sensor nodes equipped with unreplenishable and limited power resources. Since wireless communication accounts for a significant portion of node energy consumption, network lifetime and utility are dependent on the design of energy-efficient communication schemes including low-power signalling and energy-efficient multiple access protocols.

There has been a significant amount of research on transmission power control for wireless networks. Several approaches for maximizing information transmission over a shared channel subject to average power constraints have been proposed [1], [2], [3], [4], [5], [6]. [7] addresses the issue of minimizing transmission power, subject to a given amount of information being successfully transmitted and derives power control multiple access (PCMA) algorithms for autonomous channel access. [4] describes an aggregate power control scheme for a group of interfering users subject to minimal signal-to-noise (SNR) constraints. They also show that this power vector solution is Pareto-optimal since each individual nodes power is also minimized by this vector. In other words, the strategic or node-centric solution coincides with the aggregate or network-centric solution. [2], [3] then propose joint scheduling and power-control algorithms for wireless networks based on this system model. However, the above results do not explicitly account for differential delay constraints on individual nodes. In this paper, we investigate the relationship between strategic energy minimization and aggregate energy minimization when distinct delay constraints are imposed on individual nodes. Our results show that the two solutions do not always coincide (as in [4]). More surprisingly, there are certain conditions under which aggregate energy minimization is still Pareto-optimal i.e outperforms strategic behavior.

Delay is an important constraint in many wireless network applications, for example battlefield surveillance or target tracking in which data with finite lifetime-information must be delivered before a deadline. When delay constraints are introduced, the problem of energy minimization is equivalent to calculating energy-latency tradeoffs. Minimizing transmission energy subject to latency constraints has been studied recently [8], [9], [10]. [10] considers minimum energy scheduling where the total transmit power is convex in the transmission rates and a node has to finish transmitting information over a finite time horizon $T$. [11], [12] propose online modulation scaling policies for implementing energy-latency tradeoffs over a point-to-point link. Delay constraints in wireless networks can also be examined in terms of node operation under periodic duty cycles, in which time is divided into active (awake) and inactive (asleep) periods. [13], [14], [15] establish the idea of duty cycles in WSNs as a practical means of conserving node energy. In [14], nodes periodically exchange duty cycle information with each other, thereby enabling the construction of interleaved duty cycle schedules.

In this paper, we consider the problem of minimizing transmission energy for a group of users transmitting information to their receivers over a shared wireless channel over overlapping intervals or duty cycles as in the case of wireless sensor networks. Clearly, there are two approaches to transmission energy minimization. In the first approach, nodes can adjust their transmission powers to jointly minimize an aggregate transmission energy function. While this could result in higher energy depletion at some nodes[1], this approach could be beneficial in improving network lifetime, especially for wireless sensor networks containing a large number of redundant nodes (that can afford to lose energies at differential rates).

Alternatively, nodes can adjust their transmission powers strategically in response to the power behaviors of other nodes, in order to minimize their own energy consumption. A wireless network in which nodes optimize strategically could increase network utility/performance since critical nodes, for example, routing hot-spots or clusterheads/aggregator nodes in sensor networks, *might* consume less energy (depending on loads, channel quality etc.). Another reason to investigate strategic optimization is to gauge the energy outcome of selective node misbehavior. Node misbehavior can occur in insecure networks if nodes are captured by adversaries and then do not follow agreed upon transmission policies. A useful side effect of strategic energy optimization by all nodes might be to discourage selective misbehavior by some nodes, since this misbehavior will not lead to performance gains.

For the problem of energy optimization in a wireless network, it is not clear apriori (in terms of the existence of a quantitative formula) whether strategic or joint energy minimization has a bigger impact on network utility/lifetime. As seen above, a case can be made for either approach depending on the specific network and/or application. Therefore in this paper we formulate a model for evaluating strategic as well as aggregate energy minimization in delay-constrained wireless networks and compare the energy outcomes of the two approaches. For the strategic energy minimization approach, we develop a simple game-theoretic model of a 2-player power game and analytically derive conditions for the existence of Nash equilibria in this game. We then derive the power vectors for joint energy minimization and then investigate the relationship between the energy outcomes of the two approaches. We find that when duty cycles partially overlap, there exist channel and load conditions under which the energy vector for joint energy minimization is strictly lesser than the energy vector for strategic energy minimization. This is highly surprising since strategic energy optimization focuses only on reducing individual energies and should therefore benefit at least one node.

---

[1] This is because all nodes are adjusting their powers to minimize an aggregate function and not their own energies. Note that this does not occur in the system model of [4] and related work because of the Pareto-optimality of the power vectors, but could happen in the model with explicit delay constraints investigated here, as shown later.

## II. SYSTEM MODEL

In order to gain insight into the problem, we first consider a typical interference channel model consisting of two transmitter-receiver pairs: $N_{Tx_1}$ (node 1) and $N_{Tx_2}$ (node 2), transmitting to their respective receivers $N_{Rx_1}$ and $N_{Rx_2}$. Each transmitter has its own information to send to its receiver within its prescribed deadline. The transmission periods of the two nodes partially overlap. We assume that node 1 transmits first over the period $T^I = T_1 + T_2$, while node 2 starts its transmission later over the period $T^{II} = T_2 + T_3$, $T = T_1 + T_2 + T_3$, and $T^I \neq T^{II}$ in general. This can happen in wireless ad-hoc networks where different nodes initiate periodic transmissions at different times, for example in wireless sensor networks, nodes operate under duty cycles over an interval $T$ which is divided into active (awake) and inactive (asleep) periods. [14], [13]. We represent the average transmission load on each node by $\overline{R}_1 = B_1/T$ and $\overline{R}_2 = B_2/T$, where $B_1$ and $B_2$, respectively, represent the total amount of information to be transmitted by each node within its deadline. We use the notation $\mu_1$ and $\mu_2$ to represent the active ratios of the two nodes respectively, i.e. $\mu_1 = (|T_1 + T_2|/|T|)$ and $\mu_2 = (|T_2 + T_3|)/|T|$.

Let $\alpha^{(i,j)}$, $i, j \in \{1, 2\}$ be the channel attenuation factors between $N_{Tx_i}$ and $N_{Rx_j}$, which captures the effects of path-loss, shadowing and frequency nonselective fading. Our analysis focuses on the case of a slowly fading channel where the delay constraints are on the order of channel coherence time, i.e these parameters remain fixed over the active periods. It is also assumed that these channels experience independent fading. This two-user interference channel system can therefore be modeled as

$$
\begin{aligned}
r_{11}(t) &= \alpha^{(11)} s_{11}(t) + n_{11}(t),\ t \in T_1 \\
r_{12}(t) &= \alpha^{(11)} s_{12}(t) + \alpha^{(21)} s_{22}(t) + n_{12}(t),\ t \in T_2 \\
r_{22}(t) &= \alpha^{(12)} s_{12}(t) + \alpha^{(22)} s_{22}(t) + n_{22}(t),\ t \in T_2 \\
r_{23}(t) &= \alpha^{(22)} s_{23}(t) + n_{23}(t),\ t \in T_3
\end{aligned} \quad (1)
$$

where $r_{ij}(t)$ are the received baseband signals at node $N_{Rx_i}$ in the $j$th interval, $s_{ij}(t)$ are the transmit narrowband signals from node $N_{Tx_i}$ over $j$th interval with power $E\left[|s_{ij}|^2(t)\right] = P_{ij}$, and $n_{ij}(t)$ are the additive complex white Gaussian noise with power $\eta_i$. It is assumed transmitters and receivers have full access to channel state information (CSI) such that channel coding over two independent blocks by each transmit node enables error free transmissions over two periods. Single user decoding is assumed at each receiver $N_{Rx_i}$ to decode the information from its own transmit node $N_{Tx_i}$ while treating other party's information as Gaussian interference. The normalized mutual information between

$N_{Tx_i}$ and $N_{Rx_i}$ over the active periods are

$$\begin{aligned}\overline{R}_1 &= (1-\mu_2)R_{11} + (\mu_1+\mu_2-1)R_{12}\\ \overline{R}_2 &= (1-\mu_1)R_{23} + (\mu_1+\mu_2-1)R_{22}\end{aligned} \quad (2)$$

where $R_{ij} = \log_2(1+\rho_{ij})$ is the rate of node $i$ in the $j$th interval with signal-to-interference-noise-ratio (SNR) defined as follows:

$$\begin{aligned}\rho_{12} &= \frac{G^{(11)}P_{12}}{G^{(21)}P_{22}+\eta_1},\ \rho_{22} = \frac{G^{(22)}P_{22}}{G^{(12)}P_{12}+\eta_2}\\ \rho_{11} &= \frac{G^{(11)}P_{11}}{\eta_1},\ \rho_{23} = \frac{G^{(22)}P_{23}}{\eta_2}\end{aligned} \quad (3)$$

where $G^{(ij)} = |\alpha^{(ij)}|^2$.

## III. PROBLEM SETUP

We consider the problem of minimizing node transmission energies by using the system model defined above to formulate two approaches and analytically evaluate their differences: (a) strategic delay-constrained energy optimization and (b) delay-constrained aggregate energy optimization. We develop the models for the two transmitter-receiver case and analytically evaluate the differences between the two approaches.

We first model the problem of delay constrained energy minimization as a simple two player power game with the following parameters: Node 1 selects its transmit power during periods $T_1$ and $T_2$ from the space $\mathbb{P}$ of achievable transmit powers. Thus the strategy choice of node 1 is represented by $l_1 = (P_{11}, P_{12}) \in \mathbb{P} \times \mathbb{P}$. Likewise, the strategy choice of node 2 is given by $l_2 = (P_{22}, P_{23}) \in \mathbb{P} \times \mathbb{P}$. We consider only pure strategies here as opposed to the more general mixed strategy model where nodes choose their $P_{ij}$'s from a probability distribution. For notational simplicity, we define $P_{13} = P_{23} = 0$ since the nodes are not active during these time intervals.

Let $E_1$ and $E_2$ denote the transmission energy functions

$$\begin{aligned}E_1 &= T\left[P_{11}(1-\mu_2) + P_{12}(\mu_2+\mu_1-1)\right]\\ E_2 &= T\left[P_{23}(1-\mu_1) + P_{22}(\mu_2+\mu_1-1)\right]\end{aligned} \quad (4)$$

Let $R_{ij}$ represent the transmission rate of node $i$ during period $T_j$ as defined by Equation 3. Also, $\overline{R}_1 = B_1/T$ and $\overline{R}_2 = B_2/T$ represent the average transmission rates for node 1 and node 2. Then nodes 1 and 2 operate under load constraints $L_1$ and $L_2$ defined as

$$\begin{aligned}L_1 &= R_{11}(1-\mu_2) + R_{12}(\mu_2+\mu_1-1) - \bar{R}_1 = 0\\ L_2 &= R_{23}(1-\mu_1) + R_{22}(\mu_2+\mu_1-1) - \bar{R}_2 = 0\end{aligned} \quad (5)$$

Let $l = (l_i, l_{-i})$ represent a particular strategy profile of the power game. In this case, $l_{-1} = l_2$, $l_{-2} = l_1$ and $l$ also represents a particular energy outcome of the game.

We define the payoff at node $i$ under strategy profile $l$ as:

$$\Pi_i(l) = -E_i$$

Strategy $l_i$ is defined to be the best-response of player $i$ to a given $l_{-i}$ if

$$\Pi_i(l'_i, l_{-i}) \leq \Pi_i(l_i, l_{-i}) \text{ for all strategies } l'_i.$$

Let $BR_i(l_{-i})$ denote the set of player $i$'s best response to $l_{-i}$. A strategy profile $l = (l_1, l_2)$ is optimal if the nodes are playing a Nash Equilibrium[16] i.e. $l_i \in BR_i(l_{-i})$ for each sensor node $i$.

Note that the best-response power strategy of node 1 minimizes its individual energy consumption and satisfies its load constraint for a given power strategy employed by node 2, without accounting for the load constraint of the other node. However at the Nash equilibrium point, node 2 is also playing its best repones to node 1, i.e. both users are simultaneously satisfying their load constraints as well as minimizing their individual energies for each others power vector solutions. We will shortly identify system conditions (for example, load and channel quality) under which the two players arrive at Nash equilibrium in the power game, as well as the existence of Pareto-optimal (efficient) equilibria.

We also consider the joint minimization approach in which nodes jointly adjust their powers during overlapping periods in order to minimize the aggregate energy , i.e. minimize $\sum_i E_i$, subject to the load constraints $L_i$. Joint minimization is important in itself since there are circumstances under it is preferable from the application point of view, for example data aggregation in sensor networks with large number of redundant nodes. More importantly, while strategic energy optimization naturally suggests energy benefits to some nodes, we investigate whether there are conditions under which joint energy minimization can strictly dominate the strategic approach, with respect to all individual node energies. Surprisingly, this is indeed the case as shown below.

## IV. ANALYTICAL RESULTS: STRATEGIC VERSUS JOINT ENERGY MINIMIZATION

We first obtain optimal strategic power vectors followed by power vectors for joint energy minimization. Optimal strategic power vectors correspond to the Nash equilibrium points of the two player power game defined above. Generally games can have several Nash equilibria or none at all [16], depending on specific conditions (in this case channel quality and load). We analyze these possibilities and the implications below.

Let power vectors $P_1^* = (P_{11}^*, P_{12}^*)$ and $P_2^* = (P_{23}^*, P_{22}^*)$ represent node 1 and node 2's best-responses to each other in the two player power game, with $R_{12}^* =$

$\log(1+P_{12}^*/(\beta_1+\alpha_1 P_{22}^*))$ and $R_{22}^* = \log(1+P_{22}^*/(\beta_2+\alpha_2 P_{12}^*))$ the corresponding best-response rates, where $\beta_1 = \eta_1/G^{(11)}$, $\alpha_1 = G^{(21)}/G^{(11)}$, $\beta_2 = \eta_2/G^{(22)}$ and $\alpha_2 = G^{(12)}/G^{(22)}$. Also let $C_1 = 2^{\overline{R}_1/(1-\mu_2)}$ and $C_2 = 2^{\overline{R}_2/(1-\mu_1)}$ be load related terms. Finally, define $n = \mu_1/(1-\mu_2)$, $m = \mu_2/(1-\mu_1)$, $0 < \mu_1, \mu_2 < 1$, $x = 2^{R_{12}^*}$ and $y = 2^{R_{22}^*}$. Then we have,

**Proposition 1:** *The Nash equilibria of the two player power game are determined by the solutions to the system of bivariate functions $\{\mathcal{F}(R_{12}^*, R_{22}^*) = 0, \mathcal{G}(R_{12}^*, R_{22}^*) = 0\}$ defined by*

$$\begin{aligned}
\mathcal{F}: \quad & \beta_1 x^n y^m + \alpha_1 \beta_2 C_2 x^n y \\
& - \alpha_1 \beta_2 C_2 x^n - \beta_1 C_1 y^m = 0 \\
\mathcal{G}: \quad & \beta_2 x^n y^m - \beta_2 C_2 x^n \\
& + \alpha_2 \beta_1 C_1 x y^m - \alpha_2 \beta_1 C_1 y^m = 0
\end{aligned}$$

*where $x \geq 1$, $y \geq 1$.*

*Proof:* For a given $P_{22}$, the best-responses $(\tilde{P}_{11}, \tilde{P}_{12})$ of node 1 are the solutions to the constrained minimization problem

$$\text{Min } E_1 \text{ s.t } L_1 = 0$$

By the theory of Lagrange multipliers [17], this can be obtained by considering the function $H_1(\lambda_1, P_{11}, P_{12}) = E_1 - \lambda_1 L_1$, where $\lambda_1$ is a constant. The necessary condition for the local minimum of $E_1$ satisfies

$$\nabla_{\lambda_1, \tilde{P}_{11}, \tilde{P}_{12}} H_1 = \overline{0} \quad (6)$$

Solving Eq. 6 leads to

$$\frac{\lambda_1}{\ln 2} = \tilde{P}_{11} + \beta_1 \quad (7)$$

$$\frac{\lambda_1}{\ln 2} = \tilde{P}_{12} + \beta_1 + \alpha_1 P_{22} \quad (8)$$

$$\frac{\lambda_1}{\ln 2} = \frac{\beta_1 C_1}{\tilde{x}^{n-1}} \quad (9)$$

where $\tilde{x} = 1 + \frac{\tilde{P}_{12}}{\beta_1 + \alpha_1 P_{22}}$. Similarly, the best-responses $(\tilde{P}_{23}, \tilde{P}_{22})$ of node 2 for a given $P_{12}$ can be obtained from $H_2(\lambda_2, P_{23}, P_{22}) = E_2 - \lambda_2 L_2$ in an identical manner as:

$$\frac{\lambda_2}{\ln 2} = \tilde{P}_{23} + \beta_2 \quad (10)$$

$$\frac{\lambda_1}{\ln 2} = \tilde{P}_{22} + \beta_2 + \alpha_2 P_{12} \quad (11)$$

$$\frac{\lambda_1}{\ln 2} = \frac{\beta_2 C_2}{\tilde{y}^{m-1}} \quad (12)$$

$$(13)$$

where $\tilde{y} = 1 + \frac{\tilde{P}_{22}}{\beta_2 + \alpha_2 P_{12}}$.

The above equations describe the best-responses of each node to an arbitrary power value of the other node. At the Nash equilibrium point, these power values are not arbitrary and must in fact be best-responses to each other. Let $(P_{11}^*, P_{12}^*)$ and $(P_{23}^*, P_{22}^*)$ represent the Nash power vectors. They can be obtained by solving Eqs. 7-12, where all the power variables are replaced by the $P_{ij}^*$'s.

Combining Eqs. 8 and 9 and Eqs. 11 and 12, we get

$$\beta_1 + \alpha_1 P_{22}^* = \frac{\beta_1 C_1}{x^n} \quad (14)$$

$$\beta_2 + \alpha_2 P_{12}^* = \frac{\beta_2 C_2}{y^m} \quad (15)$$

Using the definitions of $x$ and $y$ and simplifying yield the equilibrium functions as stated.
∎

We now discuss under what conditions equilibria exist and if so, how many. Meaningful equilibria correspond to non-negative power allocations $P_{12}^* \geq 0$ and $P_{22}^* \geq 0$ are therefore those non negative real-valued solutions to the equilibrium functions which satisfy $1 \leq x \leq 2^{\overline{R}_1/\mu_1}$ and $1 \leq y \leq 2^{\overline{R}_2/\mu_2}$. When one of the power solutions is zero, it corresponds to TDM–Time Division Multiplexing. We now provide explicit load and channel quality conditions for the existence of Nash equilibria for the power game.

**Proposition 2:** *The power game does not have Nash equilibrium points if exactly one of condition $S$ or condition $T$ defined below is true. If $S$ and $T$ are both simultaneously true or false, then there exist at most three Nash equilibria.*

$$S:$$
$$\frac{\beta_1}{\alpha_1}(2^{\frac{\overline{R}_1}{1-\mu_2}} - 1) < \beta_2(2^{\frac{\overline{R}_2}{\mu_2}} - 1)$$
$$\frac{(n-1)\beta_2 C_2 A}{(n-1)\beta_2 A + \alpha_2 \beta_1 C_1} > \left[1 + \frac{(n-1)\beta_1(C_1 - A)}{\alpha_1(\beta_2 C_2 A(n-1) + \alpha_2 \beta_1 C_1)}\right]^m \quad (16)$$

$$T:$$
$$\frac{\beta_2}{\alpha_2}(2^{\frac{\overline{R}_2}{1-\mu_1}} - 1) < \beta_1(2^{\frac{\overline{R}_1}{\mu_1}} - 1)$$
$$\frac{(m-1)\beta_1 C_1 B}{(m-1)\beta_1 B + \alpha_1 \beta_2 C_2} > \left[1 + \frac{(m-1)\beta_2(C_2 - B)}{\alpha_2(\beta_1 C_1 B(m-1) + \alpha_1 \beta_2 C_2)}\right]^n \quad (17)$$

where $0 < \mu_1, \mu_2 < 1$.

**Corollary 1:** *For given channel quality and load conditions, there always exist duty cycle values under which the nodes can find meaningful equilibrium.*

As seen from Eqs. 16 and 17, for any channel quality and load, we can always find $\mu_1$ and $\mu_2$ such that the LSH of condition S.1 and T.1 exceed their RHS. Therefore, both $S$ and $T$ can be made false and thus equilibrium exists.

Next we identify the power vectors for the case when the two nodes carry out joint energy minimization. Let $R_{12}^J$ and $R_{22}^J$ be the rate solutions during the overlapping period and denote $x = 2^{R_{12}^J}$ and $y = 2^{R_{22}^J}$. The corresponding power solutions are denoted by $P^J$.

**Proposition 3:** *The optimal power vectors for joint energy minimization are determined by the solutions $(x, y)$ to*

$$\mathcal{P} : \beta_1 [1 + \alpha_1(x-1)] [C_1 (1 - \alpha_1\alpha_2(x-1)(y-1)) - x^n (1 + \alpha_2(y-1))] y^m$$
$$= \alpha_1\beta_2 C_2 (1 - \alpha_1\alpha_2(x-1)(y-1)) [1 + \alpha_2(y-1)] (y-1) x^n$$
$$\mathcal{Q} : \beta_2 [1 + \alpha_2(y-1)] [C_2 (1 - \alpha_1\alpha_2(x-1)(y-1)) - y^m (1 + \alpha_1(x-1))] x^n$$
$$= \alpha_2\beta_1 C_1 (1 - \alpha_1\alpha_2(x-1)(y-1)) [1 + \alpha_1(x-1)] (x-1) y^m$$

*with $P_{11}^J = \beta_1 C_1 / x^{n-1}$ and $P_{23}^J = \beta_2 C_2 / y^{m-1}$.*

*Proof:* The joint objective function to be minimized by both nodes is

$$\text{Min } (E_1 + E_2) \text{ s.t } \{L_1 = 0, L_2 = 0\}$$

Using Lagrange multiplier theory for constrained minimization [17], define the new function $D(\lambda_1, \lambda_2, P_{11}, P_{12}, P_{23}, P_{22}) = \sum_i E_i - \lambda_i L_i$, where $\lambda_i$ is the Lagrange multiplier, $i = 1, 2$. The local minimum of $\sum_i E_i$ satisfies

$$\nabla_{\lambda_1, \lambda_2, P_{11}^J, P_{12}^J, P_{23}^J, P_{22}^J} D = \overline{0} \quad (18)$$

Let $x = 1 + P_{12}^J/(\beta_1 + \alpha_1 P_{22}^J)$ and $y = 1 + P_{22}^J/(\beta_2 + \alpha_2 P_{12}^J)$. Solving Eq. 18 and simplifying leads to

$$P_{11}^J + \beta_1 = \frac{x (\beta_1 + \alpha_1 P_{22}^J)(1 + \alpha_2(y-1))}{1 - \alpha_1\alpha_2(x-1)(y-1)} \quad (19)$$

$$P_{23}^J + \beta_2 = \frac{y (\beta_2 + \alpha_2 P_{12}^J)(1 + \alpha_1(x-1))}{1 - \alpha_1\alpha_2(x-1)(y-1)} \quad (20)$$

$$P_{11}^J + \beta_1 = \frac{\beta_1 C_1}{\tilde{x}^{n-1}} \quad (21)$$

$$P_{23}^J + \beta_2 = \frac{\beta_2 C_2}{\tilde{y}^{m-1}} \quad (22)$$

Finally, using

$$x - 1 = \frac{P_{12}^J}{\beta_1 + \alpha_1 P_{22}^J} \quad (23)$$

$$y - 1 = \frac{P_{22}^J}{\beta_2 + \alpha_2 P_{12}^J} \quad (24)$$

and combining we get the results as stated. ∎

In the case of joint energy minimization, the non-existence of feasible power vectors, (i.e $P_{12}^J \geq 0$ and $P_{22}^J \geq 0$) implies that one node is creating significant interference at the other nodes receiver. In this case, (since the nodes are cooperating, unlike in the strategic optimization case) one of the nodes can choose to zero its power output during the overlapping period.

**Proposition 4:** *There exist load, channel quality and duty-cycle conditions under which joint energy minimization is Pareto-optimal, i.e. the optimal energy cost for each node under joint energy minimization is lower than its strategically optimal energy cost.*

Due to length constraints, we do not provide a formal proof and instead demonstrate the existence of such conditions through numerical examples. Figure 1 demonstrates the case when joint energy minimization is Pareto-optimal for different values of $\mu$.

## V. NUMERICAL RESULTS

This section contains numerical results for optimal power allocation given the duty cycle $\mu = \mu_1 = \mu_2$ for both the strategic and total energy minimization approach. It is assumed normalized $\beta_1 = \beta_2 = 1$ and $T = 1$.

Figure 1-3 compare individual energies $E_1$ and $E_2$, as well as the total energy $E_1 + E_2$, under both joint and strategic energy minimization schemes, respectively. It has been shown in [4] that joint energy minimization is Pareto-optimal when overlap is complete, i.e. $\mu = 1$. Figure 1 demonstrates the case when joint energy minimization is still Pareto-optimal even for partial overlap, i.e. $\mu < 1$. It can also be seen that the dominance of the joint minimization scheme over the strategic one becomes greater as $\mu$ increases. All these observations agree with the Proposition 4

Figure 2-3 reflect converging tendency of these two schemes in the sense that the difference between individual energies is decreasing. We could expect as $\mu \to 1$, joint energy minimization will yield the same energy expenditures as the strategic approach for those cases.

For intermediate $\mu$ values, Figure 3 and illustrates the benefit of the strategic approach in terms of energy gains by the user having smaller load and higher interference. Since the goal of the strategic scheme is to minimize individual energies, node 1 saves his energy at the price of higher energy consumptions for node 2 compared with the joint energy minimization scheme.

If the node with higher load has better channel quality in terms of smaller $\alpha_j$, there exists a crossing point of $\mu$ beyond which the joint minimization scheme becomes Pareto optimal as shown in Figure 2.

The figures show that both joint as well as strategic optimization have their advantages. Either scheme can be preferable depending on the specific parameters of data loads, channel qualities, and duty cycles.

## VI. CONCLUSIONS

We have formulated two versions of the power control problem for wireless networks with latency constraints arising from duty cycle allocations. In the first version, strategic power optimization, wireless nodes are modeled as rational agents who strategically adjust their powers to minimize their own energy. In the other version, joint power optimization, wireless nodes jointly minimize the aggregate energy expenditure. Using a simple two node interference channel model, we obtain insights into how strategic and joint power allocation for energy minimization is affected by data loads and channel quality. We provide analytical solutions along with numerical examples for finding strategic equilibria and also derive simple necessary conditions for the existence of such equilibria. We show that joint optimization can (sometimes) be Pareto-optimal and dominate strategic optimization, i.e the energy expenditure of all nodes is lower than if they were using strategic optimization.

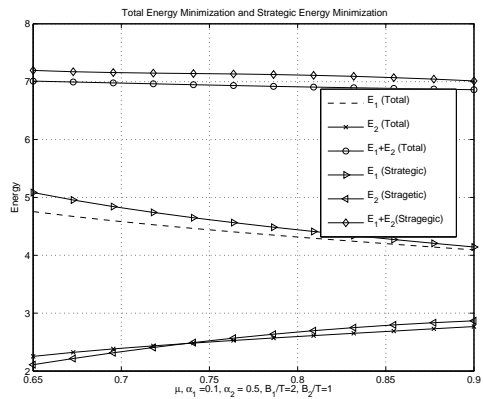

Fig. 1. Energy versus active cycle length for joint energy minimization and strategic energy minimization scheme. $B_1/T = B_2/T = 2$, $\alpha_1 = 0.2$, $\alpha_2 = 0.5$.

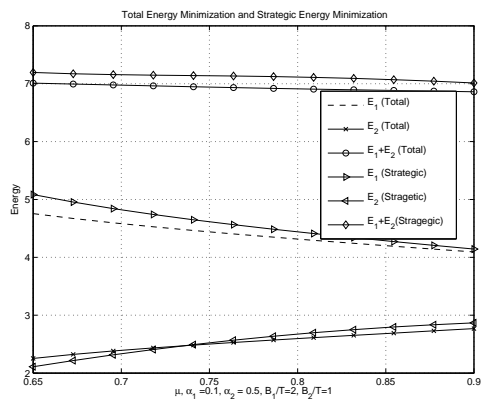

Fig. 2. Energy versus active cycle length for joint energy minimization and strategic energy minimization scheme. $B_1/T = 2$, $B_2/T = 1$, $\alpha_1 = 0.1$, $\alpha_2 = 0.5$.

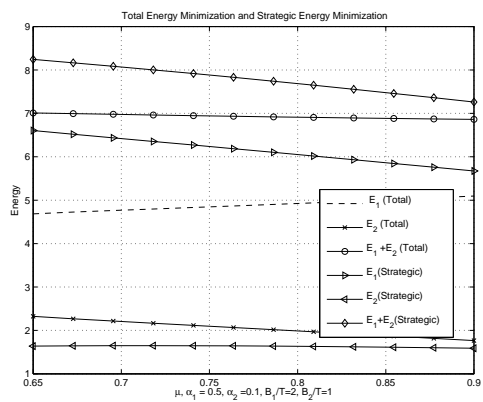

Fig. 3. Energy versus active cycle length for joint energy minimization and strategic energy minimization scheme. $B_1/T = 2$, $B_2/T = 1$, $\alpha_1 = 0.1$, $\alpha_2 = 0.5$. $B_1/T = 2$, $B_2/T = 1$, $\alpha_1 = 0.5$, $\alpha_2 = 0.1$.